\documentclass[]{preprint}
\pdfsuppressptexinfo=-1 \pdfinfoomitdate=1 \pdftrailerid{}
\usepackage[utf8]{inputenc}
\usepackage{inconsolata}
\usepackage{longtable,array,calc}

\makeatletter\@ifpackageloaded{fontawesome}{}{\providecommand{\faTags}{}}\makeatother
\hypersetup{pdftitle={Who Gets Blamed for the Blackout? A Critical Incident Analysis of Clients, Reputation Systems, and Online Freelancers},pdfauthor={Naim Bin Hasan; Fahim Bin Hasan},pdfcreator={},pdfproducer={}}
\title{Who Gets Blamed for the Blackout? A Critical Incident Analysis of Clients, Reputation Systems, and Online Freelancers}
\author[1,*]{Naim Bin Hasan}
\author[2]{Fahim Bin Hasan}
\affiliation[1]{School of Information Studies, Syracuse University, Syracuse, New York, USA}
\affiliation[2]{Dhaka College, University of Dhaka, Dhaka, Bangladesh}
\contribution[*]{Corresponding author}
\abstract{Online labor platforms such as Fiverr and Upwork judge freelancers by how fast they reply, whether they deliver on time, and what clients say about them. In the Global South, these signals depend on electricity and internet connections that workers do not control. This paper asks how infrastructure failures, such as power outages, internet failures, and government internet shutdowns, become lasting reputational penalties in the gig economy, and how freelancers prevent or repair them. We use the critical incident technique to analyze 149 incidents from 30 interviews with online freelancers in Bangladesh, including accounts of the July 2024 nationwide internet shutdown. Most breakdowns were absorbed without lasting harm. Of the 21 that became lasting penalties, 17 passed through clients. A client who could not see the power cut or the shutdown read the worker's silence as unreliability and left, canceled, or gave a bad rating. Platform reputation systems then stored that judgment without its context. We call this two-stage process client-mediated misattribution. Freelancers responded with four repair strategies: making the context visible, making the breakdown invisible, staying visible to the platform, and reducing dependence on any single platform or client. Exposure and the capacity to repair were unequal, and urban participants were the most exposed. The study extends research on algorithmic management, attribution in distributed work, and information and communication technologies for development (ICT4D), and it proposes outage-aware design for platform rating and reputation systems.}
\checkdata[\faTags\;Keywords]{online labor platforms; gig economy; reputation systems; algorithmic management; infrastructure breakdown; internet shutdown; critical incident technique; Bangladesh}
\correspondence{\email{naimbhdu@gmail.com}}
\checkdata[ORCID (Naim Bin Hasan)]{\href{https://orcid.org/0009-0002-4930-7241}{\texttt{0009-0002-4930-7241}}}
\date{November 2026}

\begin{document}
\maketitle
\newpage
\tableofcontents
\newpage
\section{Introduction}\label{introduction}

In July 2024, Meem Jahan, a 24-year-old undergraduate who designs websites for overseas clients, had six or seven projects running when the Bangladeshi government shut down the internet across the country. She lost five of those clients. The two who stayed had something the others lacked: they had seen the news about Bangladesh on the BBC. The same outage produced two different outcomes, and the difference lay in what the client could see.

Online labor platforms such as Fiverr and Upwork connect workers in low- and middle-income countries with clients abroad. Bangladesh is one of the largest suppliers of this work. In the Online Labour Index 2020, Bangladesh accounted for 15\% of the online freelancers tracked, second only to India (\textcolor{themeaccent}{Stephany et al., 2021}). These platforms present themselves as open markets that reward skill and reliability. In practice, they measure reliability through traces such as response time, on-time delivery, and client ratings. Each of those traces depends on electricity and connectivity. When the power fails or the internet goes down, the worker stops producing traces, and the people judging the worker see only an absence.

Three bodies of research help explain what happens next, but none of them explains it alone. Research on algorithmic management, meaning the use of software to direct, evaluate, and discipline workers, shows that platforms turn client ratings into lasting sanctions (\textcolor{themeaccent}{Kellogg et al., 2020; Möhlmann et al., 2021}). This work usually assumes that the worker is connected, so a missing reply reflects a choice. Research on distributed work shows that people blame remote partners for problems caused by situations they cannot see (\textcolor{themeaccent}{Cramton, 2001; Cramton et al., 2007}). It studies teams with a shared future, where a wrong impression stays private and can be corrected. Research on platform work in the Global South documents poor infrastructure among the hardships workers face (\textcolor{themeaccent}{Graham et al., 2017; Wood et al., 2019}). It does not trace how a local breakdown becomes a reputational loss.

This paper asks who is blamed when infrastructure fails, and how that blame becomes a lasting record. It addresses the gap with two research questions:

\begin{quote}
RQ1. How do local infrastructural breakdowns become lasting penalties for Global South online freelancers, and what roles do clients and platform reputation systems play?

RQ2. How do freelancers prevent or repair these penalties, and how unevenly is the capacity to do so distributed?
\end{quote}

We answer them with a critical incident analysis of 30 interviews with online freelancers in Bangladesh. The analysis yields three findings. First, breakdown is routine, and most of it is absorbed. Second, when breakdown does become a penalty, the penalty usually forms in two stages. A client who cannot see the cause attributes the silence to the worker, and the platform then records that attribution in ratings, rankings, and account decisions. We call this process client-mediated misattribution. Third, freelancers repair or prevent these penalties in four ways. Workers who served distant clients were more exposed, and only some could pay for backup power and connections.

The paper contributes to three conversations. For research on algorithmic management, it shows that a platform does not have to misread workers itself to penalize them unfairly. It can store a client's misreading and give it lasting weight. For research on attribution in distributed work, it shows what happens when a misattribution leaves the private impressions of a team and becomes a public score in a market. For research on information systems in development, it shows how platform work moves the risk of national infrastructural failure onto individual workers.

\section{Literature review}\label{literature-review}

\subsection{Platform work in the Global South}\label{platform-work-in-the-global-south}

Remote platform work offers income to workers who have few good options in their local labor markets, but on unequal terms. Graham et al. (2017) found real income gains among remote workers in Sub-Saharan Africa and Southeast Asia, together with weak bargaining power and little chance to move up. Wood et al. (2019) showed that rating-based control gave the same workers flexibility while also producing overwork, irregular hours set by clients' time zones, and sleep deprivation. Anwar and Graham (2021) describe African remote workers as caught between the freedom the work offers and the precarity it imposes. Intense global competition keeps pay low (\textcolor{themeaccent}{Beerepoot \& Lambregts, 2015}), and ratings and rankings govern access to work (\textcolor{themeaccent}{Rani \& Furrer, 2021}). Heeks (2022) calls this pattern adverse digital incorporation: inclusion in a digital system on terms that let a more advantaged group take a disproportionate share of the value.

For workers from the Global South, platform reputation is especially valuable. Agrawal et al. (2016) found on oDesk, the platform that later became Upwork, that employers were less likely to hire applicants from less developed countries, and that verified work histories raised those applicants' chances disproportionately. Lehdonvirta et al. (2019) describe platforms as signaling environments that help one-person providers in emerging economies prove their quality to foreign clients. Reputation thus offsets the doubts clients hold about workers' home countries. Anything that damages it costs these workers more than it would cost a worker whose country carries no such doubt.

Research on information systems in developing countries treats local conditions as part of how technology works, not as background. It also calls for studies of how the local and the global interact (\textcolor{themeaccent}{Avgerou, 2008; Walsham \& Sahay, 2006}). Yet the few peer-reviewed studies of Bangladeshi platform workers focus on precarity and well-being, not on the mechanics of reputation (\textcolor{themeaccent}{Billah \& Khan, 2026}). This paper examines how infrastructural failure comes to be read as worker negligence.

\subsection{Algorithmic management and customer-sourced reputation}\label{algorithmic-management-and-customer-sourced-reputation}

Kellogg et al. (2020) identify recording and rating as two of the main ways algorithms control work, and they argue that algorithmic control is more complete and more opaque than earlier forms. Möhlmann et al. (2021) show that the same platform system both matches workers to demand and controls their behavior. On a large freelance platform, Rahman (2021) found that an opaque and changing evaluation algorithm created an ``invisible cage'': workers could not tell which behaviors counted, so they experimented with the system or restricted their own conduct.

Much of this evaluation comes from customers. Fuller and Smith (1991) described how service firms used comment cards and complaint systems to turn customers into a cheap source of supervision. Platforms formalize this arrangement. Rosenblat and Stark (2016) found that Uber uses passenger ratings, which can lead to deactivation, to control drivers who cannot see how they are judged. Rosenblat et al. (2017) argue that rating systems make consumers into managers whose biases enter employment decisions without accountability. Curchod et al. (2020) found that eBay sellers experienced customer ratings, processed by the platform's algorithm, as judgments they could not contest. Cameron and Rahman (2022) show that on gig platforms, control and resistance take shape between workers and customers, not only between workers and the platform.

Research on online labor markets explains why a single rating weighs so much. Reputation decides who wins work and at what price (\textcolor{themeaccent}{Moreno \& Terwiesch, 2014}). It is hard to build from nothing, because employers offer too few first jobs to unproven workers (\textcolor{themeaccent}{Pallais, 2014}). Average ratings also drift upward over time, so most scores cluster near the top and any lower score stands out (\textcolor{themeaccent}{Filippas et al., 2022}). Reputation earned in one setting does not fully transfer to another (\textcolor{themeaccent}{Kokkodis \& Ipeirotis, 2016}), and workers experience the platform's control of their reputation as a loss of recognition (\textcolor{themeaccent}{Wood \& Lehdonvirta, 2023}).

Workers respond with anticipatory compliance, which Bucher et al. (2021) define as protecting scores in advance by accepting poor terms and over-serving clients. Sutherland et al. (2020) describe the ``gig literacies'' that freelancers build to manage clients, metrics, and reputation. In all of this work, the unfairness comes from the algorithm or from customer bias. A missing reply is treated as the worker's decision. These studies have not examined the case in which the client is simply wrong about why the worker went silent, and the platform keeps that error.

\subsection{Attribution when the situation is invisible}\label{attribution-when-the-situation-is-invisible}

Attribution theory studies how people explain other people's behavior. Observers tend to explain behavior by the person's character or effort and to underweight the situation, a tendency known as correspondence bias or the fundamental attribution error (\textcolor{themeaccent}{Gilbert \& Malone, 1995}). When observers have little information about how a person behaves across occasions, they rely on general assumptions about what usually causes such behavior (\textcolor{themeaccent}{Kelley, 1973; Kelley \& Michela, 1980}).

Cramton (2001) applied these ideas to dispersed teams. She found that team members often lacked mutual knowledge, meaning knowledge that collaborators share and know they share. They did not know about each other's local conditions, and they struggled to interpret silence. They tended to read delays as personal failings. In a later experiment, Cramton et al. (2007) showed that when a remote partner's situation is invisible, observers blame the partner more for poor performance, and that making the situation visible reduces the blame. They call this condition situation invisibility. Other studies explain why such misreadings persist. Hinds and Mortensen (2005) found that shared identity, shared context, and spontaneous communication reduce conflict in distributed teams. Trust in global virtual teams depends on timely, predictable replies (\textcolor{themeaccent}{Jarvenpaa \& Leidner, 1999}). In text-based communication, people read a great deal into the few cues they have, including timing (\textcolor{themeaccent}{Walther, 1996}). Constant connectivity has also raised expectations of immediate response (\textcolor{themeaccent}{Mazmanian et al., 2013}).

This research predicts that distant clients will misread a freelancer's silence. It was built, however, on studies of teams. Team members share a project and a future, and they have time to correct an impression. A client on a freelance platform often has neither. The hidden situations in team studies were workloads and schedules, not a national power grid or a state-ordered shutdown. Most important, a team member's misattribution stays private. On a platform, it can become a public score.

\subsection{Infrastructure, breakdown, and repair}\label{infrastructure-breakdown-and-repair}

Star and Ruhleder (1996) argued that infrastructure is a relation rather than a thing. It is infrastructure for someone, within a practice, and it becomes visible when it breaks down. Star (1999) urged researchers to study the invisible work that keeps infrastructure running. Larkin (2013) challenged the idea that infrastructure is invisible until it fails. Drawing on Nigeria, he showed that in many Global South settings breakdown is routine and infrastructure is highly visible. Graham and Thrift (2007) argue that failure is the normal state of networked systems, which constant and mostly unseen repair keeps running. In information systems research, Alter (2014) defines workarounds as goal-driven adaptations that get around obstacles in a work system, and Azad and King (2012) show how workarounds become shared routines.

Unreliable infrastructure has measurable costs. In India, electricity shortages reduce industrial output, and backup generation is an expensive coping strategy that falls hardest on small producers (\textcolor{themeaccent}{Allcott et al., 2016}). Internet shutdowns are a deliberate form of breakdown. Freyburg and Garbe (2018) link election-time shutdowns in sub-Saharan Africa to state control of internet providers, and Wagner (2018) describes Pakistan's repeated shutdowns as ruptures in communication.

This literature asks how infrastructure becomes visible in breakdown. It rarely asks to whom. Studies of repair usually examine people who share the context of a failure, and studies of shutdowns usually examine states and citizens. Platform freelancing separates the person who experiences a breakdown from the person who judges its effects. The only link between them is the infrastructure that has failed.

\subsection{Analytical frame}\label{analytical-frame}

Read together, these literatures suggest a sequence that none of them describes alone. A local breakdown interrupts a worker's activity. The client sees only silence and, without knowledge of the cause, attributes it to the worker. The platform records that attribution as a rating, a cancellation, or a response metric, and the features of reputation systems described above make the record last. Because reputation is what offsets clients' doubts about Global South workers, the loss falls hardest on them. We used this sequence as a set of sensitizing ideas during analysis, not as hypotheses to test.

\section{Research context}\label{research-context}

Bangladesh's freelancers work in conditions of frequent infrastructural failure. Participants in this study described scheduled power cuts known locally as load-shedding, cable faults at local internet providers, and connections that failed during storms. The most severe failure was political. During protests over quotas in civil-service jobs, the government cut mobile internet and then broadband across the country. Network measurements place the full blackout from 18 to 23 July 2024. Mobile data returned on 28 July, and 4G service was cut again on 4 August (\textcolor{themeaccent}{Belson, 2024; Internet Society Pulse, 2024}). A United Nations fact-finding mission concluded that the shutdowns were timed to obstruct protest organizing (\textcolor{themeaccent}{Office of the United Nations High Commissioner for Human Rights, 2025}). Bangladeshi media reported that freelancers lost contracts and long-term clients during the blackout, and that one Fiverr seller's Top Rated status was downgraded for six months (\textcolor{themeaccent}{Billah \& Hossain, 2024}).

The two platforms in this study turn responsiveness into public, durable records. Fiverr assigns freelancers to levels using metrics that include rating, response rate, and a Success Score, and it shows the level to clients (\textcolor{themeaccent}{Fiverr, n.d.-d}). Response rate is the share of first replies sent within 24 hours over the previous 90 days, and the average response time appears on the freelancer's profile (\textcolor{themeaccent}{Fiverr, n.d.-a}). The Success Score covers communication, on-time delivery, and cancellations. Its help page states that unresponsive clients do not lower the freelancer's score, but it makes no similar allowance for the freelancer's own outages (\textcolor{themeaccent}{Fiverr, n.d.-c}). Freelancers can switch on an availability setting in advance, but active orders must still be delivered on time (\textcolor{themeaccent}{Fiverr, n.d.-b}). Upwork's Job Success Score is calculated from contract outcomes and client feedback, and Upwork warns that freelancers who score below 79\% may struggle to reach new clients (\textcolor{themeaccent}{Upwork, n.d.-a, n.d.-b}). These rules assume that the worker can always be reached.

\section{Method}\label{method}

\subsection{Data}\label{data}

This study uses 30 semi-structured interviews, conducted in Bangla over Zoom and Google Meet between May and August 2025. Each interview lasted 45 to 70 minutes. Participants were recruited through personal networks, through referrals, and through Bangladeshi freelancer groups on Facebook. To take part, a person had to be at least 18, be a Bangladeshi national, have an active profile on at least one freelance platform, and have completed at least three paid projects worth more than US\$300 in total. The sample includes 21 men and 9 women, mostly in their twenties, working in graphic design, web and app development, digital marketing, content writing, and video work. Twenty-five were students. \hyperref[tab:1]{Table~1} summarizes the sample.

{\small
\begin{longtable}[]{@{}
  >{\raggedright\arraybackslash}p{(\linewidth - 10\tabcolsep) * \real{0.0528}}
  >{\raggedright\arraybackslash}p{(\linewidth - 10\tabcolsep) * \real{0.2406}}
  >{\raggedright\arraybackslash}p{(\linewidth - 10\tabcolsep) * \real{0.0869}}
  >{\raggedright\arraybackslash}p{(\linewidth - 10\tabcolsep) * \real{0.0637}}
  >{\raggedright\arraybackslash}p{(\linewidth - 10\tabcolsep) * \real{0.3483}}
  >{\raggedright\arraybackslash}p{(\linewidth - 10\tabcolsep) * \real{0.2078}}@{}}
\caption{Participants}\label{tab:1}\\

\toprule\noalign{}
\begin{minipage}[b]{\linewidth}\raggedright
\textbf{\#}
\end{minipage} & \begin{minipage}[b]{\linewidth}\raggedright
\textbf{Pseudonym}
\end{minipage} & \begin{minipage}[b]{\linewidth}\raggedright
\textbf{Gender}
\end{minipage} & \begin{minipage}[b]{\linewidth}\raggedright
\textbf{Age}
\end{minipage} & \begin{minipage}[b]{\linewidth}\raggedright
\textbf{Main work}
\end{minipage} & \begin{minipage}[b]{\linewidth}\raggedright
\textbf{Location type*}
\end{minipage} \\
\midrule\noalign{}
\endhead
\bottomrule\noalign{}
\endlastfoot
1 & Arnob Khan & M & 28 & Graphic Design & Village \\
2 & Tanvir Ahmed & M & 27 & Web Development & Unclear \\
3 & Ashfaqur Rahman & M & 23 & Digital Marketing & Village \\
4 & Faisal Chowdhury & M & 23 & Web Development & Urban \\
5 & Israt Jahan & F & 30 & SEO, Social Media Marketing & District town \\
6 & Sanjida Rahman & F & 21 & SEO, Domain Marketing & Urban \\
7 & Junaid Hasan & M & 25 & Web Development, Graphic Design & District town \\
8 & Lamia Siddiqua & F & 26 & Graphic Design & Urban \\
9 & Afsana Islam & F & 23 & Graphic Design, Web Design & Urban \\
10 & Rezwanul Haque & M & 24 & Graphic Design, Content Writing & Urban \\
11 & Mehedi Hasan Limon & M & 26 & Digital Marketing, Photography, Graphic Design & Urban \\
12 & Tariqul Islam & M & 21 & Graphic Design & District town \\
13 & Emon Shikder & M & 21 & Digital Marketing, Voice over & District town \\
14 & Sirajul Islam & M & 25 & Digital Marketing, App Review & District town \\
15 & Arif Hossain & M & 21 & App Development & District town \\
16 & Shapna Islam & F & 20 & Graphic Design & District town \\
17 & Imran Mahmud & M & 21 & App Development, Digital Marketing & District town \\
18 & Minhazul Abedin & M & 20 & Graphic Design & District town \\
19 & Sajjad Hossain & M & 24 & Academic writing, Graphic Design & Urban \\
20 & Wasim Akram & M & 23 & Graphic Design & Urban \\
21 & Mostafizur Rahman & M & 25 & Graphic Design, Video Editing & Urban \\
22 & Enamul Haque & M & 25 & Web Development & Urban \\
23 & Jakir Khan & M & 25 & VFX, Video Editing, Digital Marketing, Web Development & Urban \\
24 & Moniruzzaman Khan & M & 25 & Web Development & Village \\
25 & Jahidul Islam & M & 24 & Graphic Design, Motion Graphics & Urban \\
26 & Meem Jahan & F & 24 & Web Design, Machine Learning & Urban \\
27 & Kazi Abdul Karim & M & n/a & App Development & Village \\
28 & Nusrat Fatema & F & 23 & Digital Marketing & District town \\
29 & Humayra Anjum & F & 23 & Content Writing, Profile Setup, Graphic Design & Urban \\
30 & Sumaiya Chowdhury & F & 20 & Various website/posting tasks & District town \\
\end{longtable}
}

{\small *Location type was inferred from what participants said in the interviews. Gender and age are taken from the participant records.\par}

The interviews were collected for a broader study of freelancing and precarity. The interview guide asked about technical problems, power cuts, payments, ratings, platform experiences, and client relationships, but it did not ask specifically about how outages affected ratings or rankings. This paper is therefore a secondary analysis with a new question and a new analytic method. The study was approved by the Institutional Review Board of Florida Atlantic University (protocol IRB2502046). All names are pseudonyms.

We also used the public help-center documentation of Fiverr and Upwork to identify the metrics through which client judgments become records (Section 3).

\subsection{Critical incident analysis}\label{critical-incident-analysis}

We used the critical incident technique, which asks participants to describe concrete events that matter for an outcome of interest and then classifies those events (\textcolor{themeaccent}{Flanagan, 1954}). The technique suits interview data that were not collected for a single question, because it treats each described event as the unit of analysis rather than the interview (\textcolor{themeaccent}{Butterfield et al., 2005}). We defined an incident as a concrete episode, or a clearly described recurring pattern, in which a disruption affected a participant's platform work or in which a client or platform penalized the participant. We also recorded general beliefs about such events, but we kept them separate and gave them less weight.

We coded each incident on six dimensions. The trigger is what went wrong first, such as a power cut, an internet failure, the July 2024 shutdown, illness, or an opaque platform decision. The channel is the route by which the incident became a cost, such as a client leaving, a bad rating, a fall in ranking, or an account restriction. We also coded the actor who imposed the cost, the severity of the loss, whether the participant recovered, and any repair actions. Each incident is linked to a verbatim Bangla excerpt and its line number in the transcript.

The first extraction pass was carried out with the help of a large language model, Claude (Anthropic), which read each transcript in full and proposed incidents in a fixed format. A script then confirmed that every Bangla excerpt appeared word for word at the cited lines. We then read every proposed incident against the transcript and checked its codes. This verification confirmed all 149 incidents and their codes, so no incident was added or changed.

To check the reliability of the coding, we drew a random sample of 30 incidents (20\% of the total). A language model from a different developer, a GPT model (OpenAI), coded the trigger and channel of each sampled incident from the transcript passage alone, without access to our codes. We compared its codes with our verified codes using Cohen's kappa, a measure of agreement between two coders that corrects for agreement expected by chance (\textcolor{themeaccent}{Cohen, 1960}). Agreement was 0.88 for trigger and 0.76 for channel. By a widely used benchmark, these values indicate almost perfect and substantial agreement, respectively (\textcolor{themeaccent}{Landis \& Koch, 1977}). The disagreements involved seven of the 30 incidents. Most concerned incidents that fit two categories, such as newcomer barriers that both lowered a profile's visibility and cost the worker clients. We reviewed each disagreement against the transcript and kept the verified code. None of the disagreements changes the number of breakdown incidents that became penalties or the channels through which they did so. All interpretive decisions are our own. The quotations were translated from Bangla.

\subsection{Cross-case comparison}\label{cross-case-comparison}

To address the second research question, we built a matrix with one row per participant. It records the participant's location type, which we inferred from the interview as an urban center, a district town, or a village. It also records whether the participant had backup power or a second internet connection, whether they found clients outside the platforms, and whether any breakdown had led to a lasting penalty. We used this matrix to compare cases, not to test associations statistically. With 30 participants, the counts describe patterns in this sample only.

\subsection{Analysis and positionality}\label{analysis-and-positionality}

We followed the principles of interpretive field research in information systems, moving between individual incidents and the emerging account of the whole (\textcolor{themeaccent}{Klein \& Myers, 1999; Walsham, 2006}). We report counts to show how widespread a pattern was and quotations to show how participants understood it. We sought out cases that ran against the emerging account, such as breakdowns that caused no harm.

One of us worked as a freelance web developer on Upwork and Fiverr for more than three years before this study. That experience helped with recruiting participants and understanding platform terms without long explanations. It also created a risk that this experience would shape how we read the participants' accounts. To limit that risk, we grounded every claim in the incident table and returned to the Bangla transcripts whenever a code depended on interpretation.

\section{Findings}\label{findings}

\begin{figure}[htbp]
\centering
\includegraphics[width=\linewidth,height=0.75\textheight,keepaspectratio]{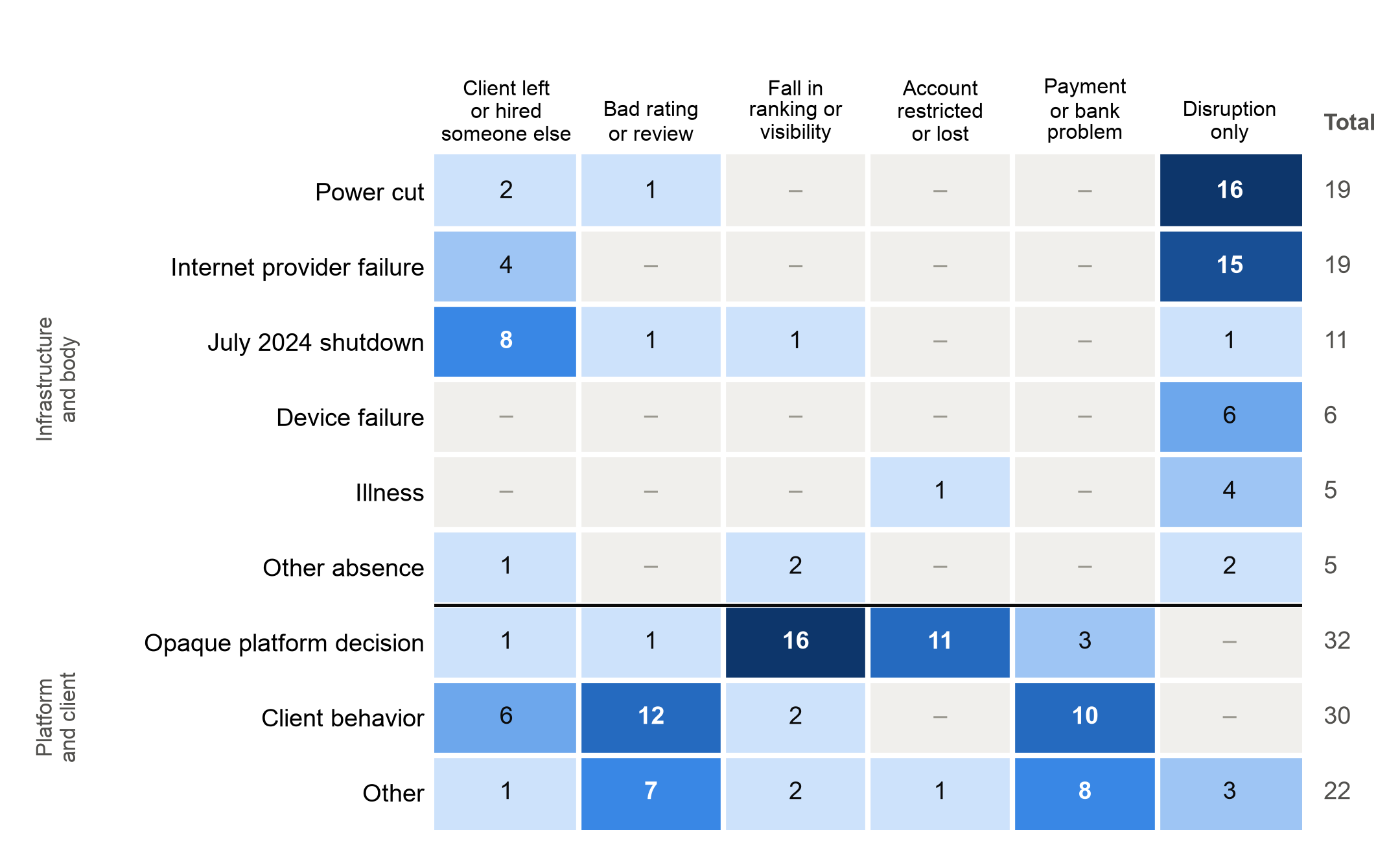}
\caption{Incidents by trigger and channel (n = 149). Darker cells hold more incidents. Rows above the line are breakdowns in infrastructure or the body.}\label{fig:1}
\end{figure}

The analysis identified 149 incidents across the 30 interviews. Of these, 41 were specific episodes, 70 were recurring patterns, and 38 were general beliefs. Sixty-five incidents began with a breakdown in infrastructure or in the worker's body, and 27 of the 30 participants reported at least one. \hyperref[fig:1]{Figure~1} shows how these incidents were distributed across triggers and channels.

\subsection{Breakdown is routine, and most of it is absorbed}\label{breakdown-is-routine-and-most-of-it-is-absorbed}

Breakdown was an ordinary part of freelance work for 27 of the 30 participants. Power cuts and internet failures each accounted for 19 incidents, the July 2024 shutdown for 11, device failures for 6, and illness for 5. Sajjad Hossain, a 24-year-old undergraduate who writes academic content and designs graphics, kept two internet connections because one of them failed so often:

\begin{quote}
``Trust me, almost 10 days out of 30 in a month I had to switch lines, because back then there were local operators.''
\end{quote}

Most breakdowns did not become lasting penalties. Of the 65 breakdown incidents, 44 caused disruption without any loss that the participant reported. Participants renegotiated deadlines, redid lost files, or switched to mobile data. Arnob Khan, a 24-year-old sociology graduate who has worked as a full-time graphic designer for five to six years, described how clients responded when load-shedding threatened a deadline:

\begin{quote}
``And when the power went out, or it looked like the deadline would pass but I couldn't deliver, there was a compromise, a negotiation with them. They took it well.''
\end{quote}

Moniruzzaman Khan is a 25-year-old undergraduate web developer whose village home lost power about every hour. He missed deadlines but kept his clients by telling them what had happened: ``I had to deliver after the deadline, had to tell them there's no power, I'll send your file a little later.'' He then bought mobile data to send the files from his laptop.

These cases show that a breakdown does not automatically become a penalty. The question is when it does. The incidents point to two conditions. First, the breakdown struck at a moment when timing mattered, such as a new inquiry, a delivery deadline, or a live call. Second, the client could not see what had happened.

\subsection{Stage one: the client misreads the silence}\label{stage-one-the-client-misreads-the-silence}

\begin{figure}[htbp]
\centering
\includegraphics[width=\linewidth,height=0.75\textheight,keepaspectratio]{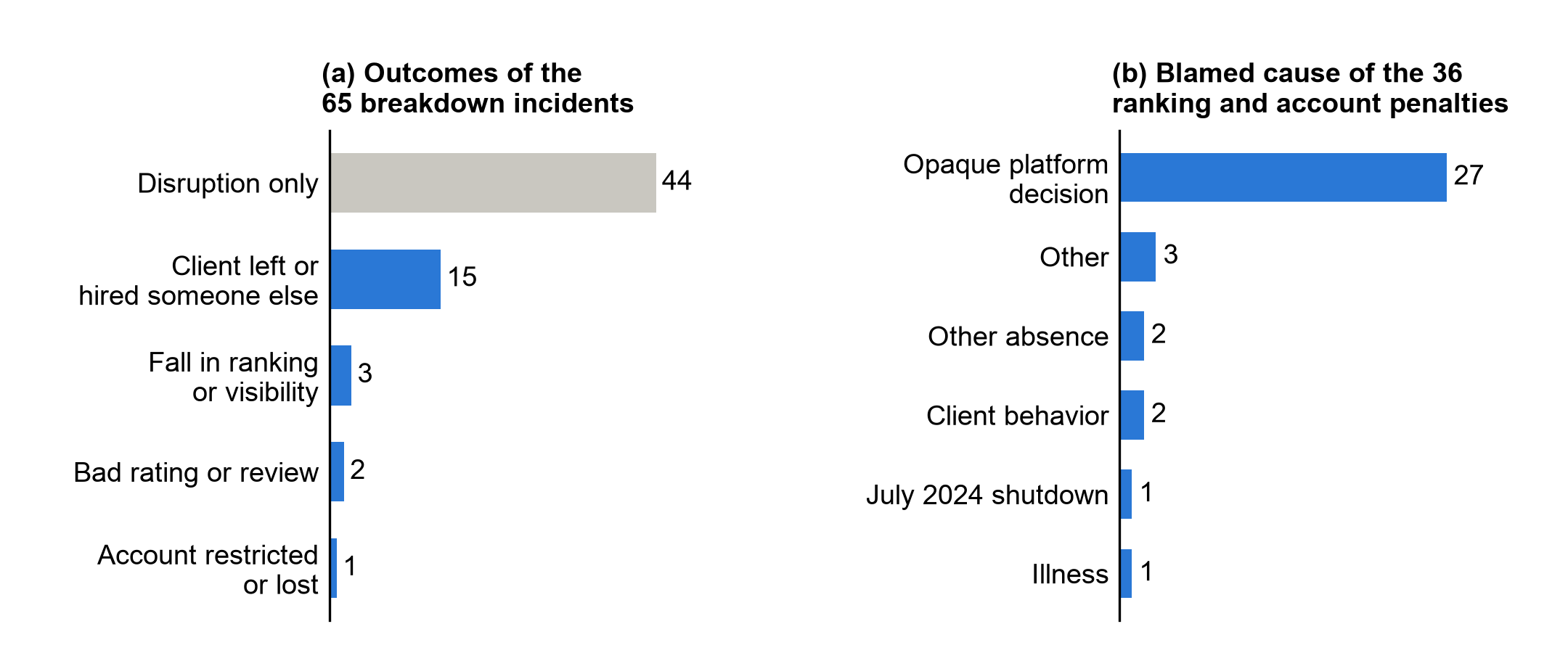}
\caption{(a) Outcomes of the 65 breakdown incidents. Gray marks breakdowns that caused disruption only. (b) What participants blamed for the 36 incidents involving a fall in ranking or an account restriction.}\label{fig:2}
\end{figure}

Twenty-one breakdown incidents, reported by 13 participants, became lasting costs. In 17 of them, the cost came through the client. In 15 incidents the client left or hired someone else, and in 2 the client left a bad rating. Only 4 ran through the platform's ranking or account systems. \hyperref[fig:2]{Figure~2a} shows where all 65 breakdown incidents ended up.

The most common pattern involved a new inquiry. Israt Jahan, a 30-year-old MBA graduate who does search engine optimization and social media marketing while caring for her paralyzed mother, lost a prospective client to a power cut:

\begin{quote}
``Because at that moment the power went out. When I replied half an hour or an hour later, the client had hired someone else.''
\end{quote}

Lamia Siddiqua, a 26-year-old undergraduate graphic designer in Dhaka, described the same sequence. Her internet provider switched off service during lightning: ``while I was talking with the client the net went off, and when I came back the client had got it from someone else.'' Mehedi Hasan Limon, a 26-year-old digital marketer, said that slow connections had cost him clients in the same way. In none of these accounts did the client wait for an explanation.

One participant believed that his client had read the silence as dishonesty. Junaid Hasan, a 25-year-old electrical engineering graduate who has done web development and graphic design since 2020, delivered a project on the final day before the deadline. Load-shedding and an internet failure then cut him off from the client at the last moment:

\begin{quote}
``I don't know what he thought, but it may have seemed to him that I was lying. So he gave me a bad review.''
\end{quote}

Junaid had no means of showing the client that the power had gone out. His explanation arrived after the fact and looked like an excuse. This is the core of the first stage. The client did not judge the breakdown, because the client never saw it. The client judged the silence and filled in its cause.

The July 2024 shutdown turned this individual problem into a collective one. Of the eight participants who discussed the shutdown, six reported losing clients, projects, or a rating. Wasim Akram, a 23-year-old full-time graphic designer who has freelanced since 2018, said: ``In July when the internet went, I found all my clients had left.'' Moniruzzaman Khan had to start over: ``Because the net went off, many of my clients left. It was quite hard for me then, building up again from zero.'' Enamul Haque, a 25-year-old undergraduate web developer, lost the projects he had in hand and received a bad review during the same period. Sajjad Hossain explained why clients did not wait:

\begin{quote}
``Even if you leave, he has 100 or 200 more like you from around the world. So even if you explain nicely that your PC broke, your laptop broke, you have no internet connection, there's no electricity at home, or a movement like the July movement happened in your country and you were fully blacked out, your buyer will never consider these things.''
\end{quote}

The contrast cases are the most revealing. In several incidents, a client did learn the cause of the silence, and the outcome changed. Meem Jahan, whose case opens this paper, kept the two clients who had seen the news:

\begin{quote}
``Two of them knew the situation in the country. They had seen the news on BBC, so they considered me. But the other five just said sorry and gave the project to someone else.''
\end{quote}

Wasim Akram kept two or three Bangladeshi clients during the shutdown because he could phone them, while his foreign clients were unreachable: ``abroad you can't contact directly by phone without internet.'' Mehedi Hasan Limon described the problem from the other side: his clients in the United States ``don't understand that our country has a system of shutting {[}the internet{]} off.'' In each case, what decided the outcome was not the breakdown itself but whether the client could see it. This matches the experimental finding that observers blame remote partners less when the partner's situation is visible (\textcolor{themeaccent}{Cramton et al., 2007}). The difference here is that the hidden situation was a national infrastructure, and the relationship was a market exchange that the client could end at once.

\subsection{Stage two: the platform turns the judgment into a record}\label{stage-two-the-platform-turns-the-judgment-into-a-record}

The second stage begins when the platform records the client's judgment. Participants described ratings as the main link between a client's opinion and their future visibility. Rezwanul Haque, a 24-year-old undergraduate graphic designer and content writer, said that ratings directly shaped ``the ranking of the profile and the gigs.'' Enamul Haque said that if a client gives even a slightly bad review, ``it has a huge impact on the freelancer.'' Junaid Hasan explained that two bad reviews among many good ones had a drastic effect on his profile. These accounts fit the finding that inflated rating systems make any lower score stand out (\textcolor{themeaccent}{Filippas et al., 2022}).

The platform also read absence directly. Mehedi Hasan Limon described his understanding of how Fiverr treats inactivity:

\begin{quote}
``If you are inactive for a long time Fiverr may think you have left the platform, so algorithm-wise your ranking will drop far down.''
\end{quote}

For Sanjida Rahman, a 21-year-old law student who has done search engine optimization and domain marketing since 2020, absence caused by illness ended a well-established account:

\begin{quote}
``If you stay away from freelancing for a long time (1-2 months), your Fiverr profile goes down a lot. I could not open my Fiverr account for about a month due to being hospitalized. Due to that, my profile got completely dismissed.''
\end{quote}

Sanjida's case shows the second stage without the first. No client misread her situation. The platform's own rules treated a month in hospital the same way they would treat a decision to quit. Her case was rare, however. Only 4 of the 21 breakdown penalties were linked by participants directly to the platform's ranking or account systems.

There is a second reason the platform's role appears small in these accounts. Participants rarely knew why a platform penalty had happened. Of the 36 incidents in which a participant described a fall in ranking or an account restriction, 27 were attributed to opaque platform rules rather than to any specific event (\hyperref[fig:2]{Figure~2b}). Sajjad Hossain said his Fiverr profile had been banned several times and called it ``a Bangladesh issue, many people's profiles get banned here.'' This opacity matches Rahman's (2021) account of the invisible cage. It also means that a freelancer who lost ranking after an outage might never connect the two. The counts in this study therefore probably understate the platform's role.

Penalties were also slow to undo. Eight of the 21 breakdown penalties were never recovered, and only 2 were fully recovered. Rebuilding was hard because new and damaged profiles struggle to attract clients. Several participants described months without orders after starting out, and Imran Mahmud, a 21-year-old diploma student who develops apps, said that when a profile disappears, ``he doesn't just lose his account. He actually loses all his clients.'' This is consistent with evidence that employers offer too few first jobs to workers without a track record (\textcolor{themeaccent}{Pallais, 2014}).

\subsection{Repair: four strategies}\label{repair-four-strategies}

\begin{figure}[htbp]
\centering
\includegraphics[width=\linewidth,height=0.75\textheight,keepaspectratio]{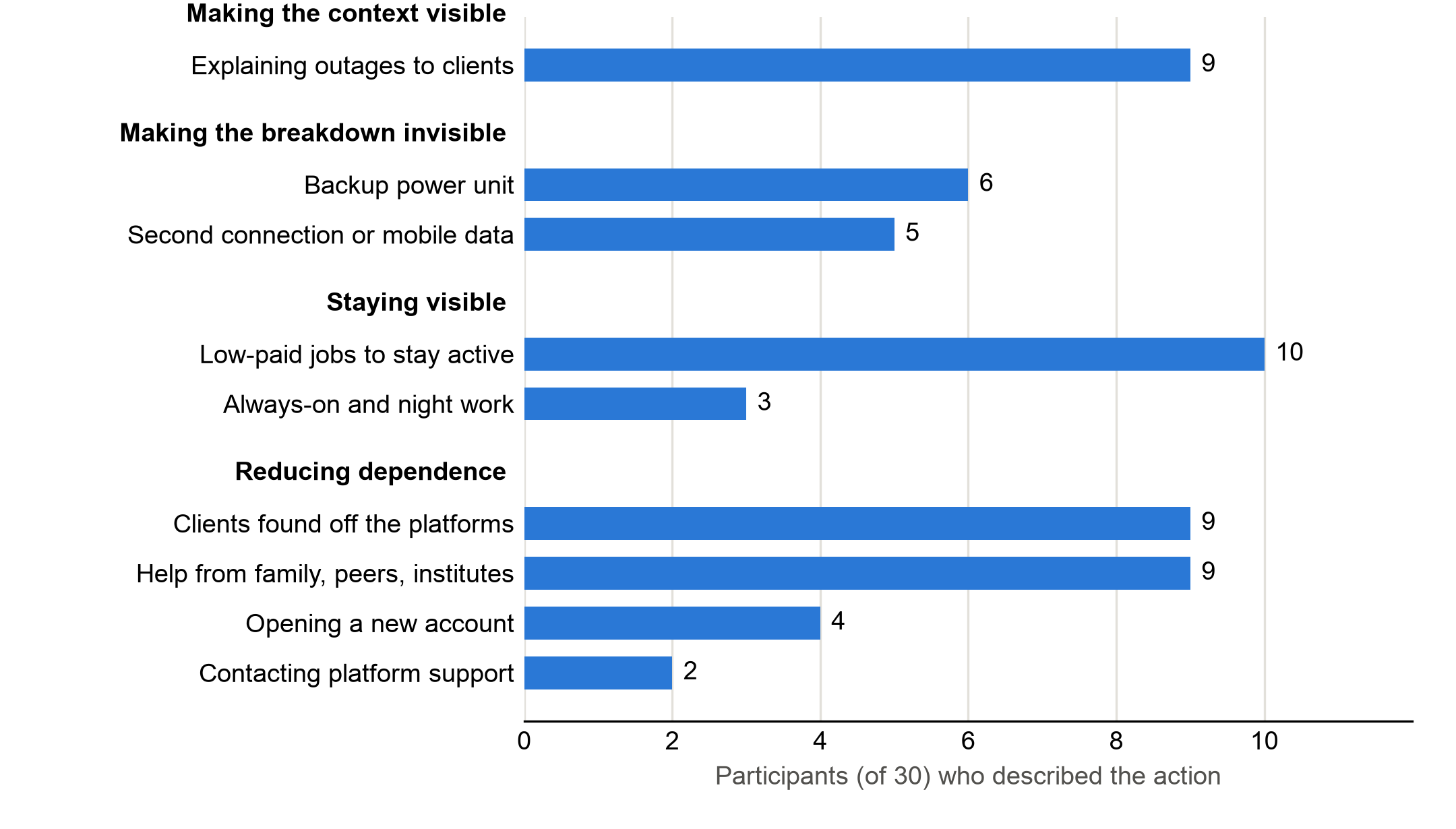}
\caption{Repair actions by strategy. Bars show the number of participants (out of 30) who described each action; one participant can describe several.}\label{fig:3}
\end{figure}

Freelancers acted both before and after breakdowns. Their actions targeted different points in the two-stage process. \hyperref[tab:2]{Table~2} groups these actions into four strategies and shows what each strategy required. \hyperref[fig:3]{Figure~3} shows how many participants described each action.

{\small
\begin{longtable}[]{@{}
  >{\raggedright\arraybackslash}p{(\linewidth - 6\tabcolsep) * \real{0.2244}}
  >{\raggedright\arraybackslash}p{(\linewidth - 6\tabcolsep) * \real{0.3739}}
  >{\raggedright\arraybackslash}p{(\linewidth - 6\tabcolsep) * \real{0.1816}}
  >{\raggedright\arraybackslash}p{(\linewidth - 6\tabcolsep) * \real{0.2201}}@{}}
\caption{Repair strategies}\label{tab:2}\\

\toprule\noalign{}
\begin{minipage}[b]{\linewidth}\raggedright
\textbf{Strategy}
\end{minipage} & \begin{minipage}[b]{\linewidth}\raggedright
\textbf{Actions described}
\end{minipage} & \begin{minipage}[b]{\linewidth}\raggedright
\textbf{Stage it targets}
\end{minipage} & \begin{minipage}[b]{\linewidth}\raggedright
\textbf{What it requires}
\end{minipage} \\
\midrule\noalign{}
\endhead
\bottomrule\noalign{}
\endlastfoot
Making the context visible & Explaining outages to clients; phoning clients; traveling to a place with a signal to send updates & Stage one (client misattribution) & A working channel to the client and a client who believes the explanation \\
Making the breakdown invisible & Battery backup units; a second internet connection; switching to mobile data & Stage one & Money for equipment and data \\
Staying visible & Taking low-paid jobs to stay active or earn ratings; working at night across time zones; refreshing profiles & Stage two (platform record) & Time, sleep, and accepting lower pay \\
Reducing dependence & Finding clients off the platforms; help from family, peers, or institutes; opening new accounts & Both stages & Social networks and skills to sell outside the platforms \\
\end{longtable}
}

The first strategy was to make the context visible to the client, which addresses the first stage directly. Nine participants described explaining outages to clients. Meem Jahan traveled to a place with a signal during a family visit so that she could send updates: ``I'd go somewhere a bit far, say around two in the afternoon, and then message the client: sir, I'm working on it.'' This strategy worked when the client could be reached and believed the explanation. It failed during the shutdown, when foreign clients could not be reached at all.

The second strategy was to make the breakdown invisible, so that the client never noticed it. Six participants described backup power units, and five described a second internet connection or mobile data. Arif Hossain, a 21-year-old diploma student who develops apps, installed a battery backup unit, known locally as an IPS, at home: ``even if the power goes, I can still communicate with clients.'' Rezwanul Haque bought one after learning how disconnection looked to his clients: ``when I understood that for foreigners it's very rude to suddenly get disconnected, I planned and started using one.'' In effect, these workers paid to supply themselves with the reliable infrastructure that the platform's rules assumed. This resembles the costly self-generation of power that Allcott et al. (2016) describe among small Indian firms.

The third strategy was to stay visible to the platform. Ten participants described taking low-paid work to keep their profiles active or to earn ratings. Jahidul Islam, a 24-year-old graduate in graphic design and motion graphics, said: ``In the beginning, I undertook several projects for only \$5 to \$10 in order to raise my profile's visibility and rating.'' Others described working through the night across time zones. These actions match the anticipatory compliance described by Bucher et al. (2021), with one difference: here, workers were also protecting themselves against breakdowns they could not predict.

The fourth strategy was to reduce dependence on any single platform or client. Nine participants found clients through Facebook, LinkedIn, or referrals, and nine relied on family, peers, or training institutes to recover accounts or solve problems. Ashfaqur Rahman is a 23-year-old undergraduate who runs a small digital marketing team. He kept working through village power cuts because his teammates in Dhaka continued the work: ``I'm in the village but all my work is running, because I have teammates who are working while I'm home.'' Moving work off the platform reduced exposure to ratings, but it also gave up the payment protection that the platforms provide.

\subsection{Unequal exposure and unequal capacity}\label{unequal-exposure-and-unequal-capacity}

Exposure to client-mediated misattribution was not evenly spread. \hyperref[tab:3]{Table~3} compares participants by location and by access to backup equipment. Ten of the 14 urban participants had at least one breakdown that became a lasting penalty, compared with 2 of the 11 participants in district towns and 1 of the 4 in villages.

{\small
\begin{longtable}[]{@{}
  >{\raggedright\arraybackslash}p{(\linewidth - 4\tabcolsep) * \real{0.4487}}
  >{\raggedright\arraybackslash}p{(\linewidth - 4\tabcolsep) * \real{0.2137}}
  >{\raggedright\arraybackslash}p{(\linewidth - 4\tabcolsep) * \real{0.3376}}@{}}
\caption{Breakdown penalties by location and backup equipment. Counts describe this sample only.}\label{tab:3}\\

\toprule\noalign{}
\begin{minipage}[b]{\linewidth}\raggedright
\textbf{Group}
\end{minipage} & \begin{minipage}[b]{\linewidth}\raggedright
\textbf{Participants}
\end{minipage} & \begin{minipage}[b]{\linewidth}\raggedright
\textbf{At least one breakdown penalty}
\end{minipage} \\
\midrule\noalign{}
\endhead
\bottomrule\noalign{}
\endlastfoot
By location (inferred) & & \\
Urban center & 14 & 10 (71\%) \\
District town & 11 & 2 (18\%) \\
Village & 4 & 1 (25\%) \\
Unclear & 1 & 0 (0\%) \\
By backup equipment & & \\
Backup power or a second connection & 9 & 5 (56\%) \\
Neither & 21 & 8 (38\%) \\
All participants & 30 & 13 (43\%) \\
\end{longtable}
}

This is the opposite of what infrastructure quality alone would predict, since rural participants described worse power and connectivity. One possible explanation is exposure. Urban participants may have served more international clients and held more running projects, so they had more to lose when a client could not see the cause of a silence. Some district-town participants worked mainly for local clients, who could be reached by phone and knew local conditions. If this explanation holds, exposure depends on the distance between worker and client as much as on the quality of the infrastructure.

The capacity to repair was also unequal. Backup power and a second connection cost money, and several participants bought them only after a loss. Among the nine participants who had backup power or a second connection, five had experienced a breakdown penalty, compared with eight of the 21 who had neither. This does not show that backup equipment fails to help. It reflects the order of events: Rezwanul Haque and Jahidul Islam both bought backup equipment after costly disconnections. Participants without equipment described different limits. Moniruzzaman Khan had no backup power at his village home when it lost power every hour. Others worked only from phones. Rural participants thus faced worse infrastructure with fewer private remedies, and what protected some of them was a client base close enough to understand local conditions.

\section{Discussion}\label{discussion}

\subsection{Client-mediated misattribution}\label{client-mediated-misattribution}

\begin{figure}[htbp]
\centering
\includegraphics[width=\linewidth,height=0.75\textheight,keepaspectratio]{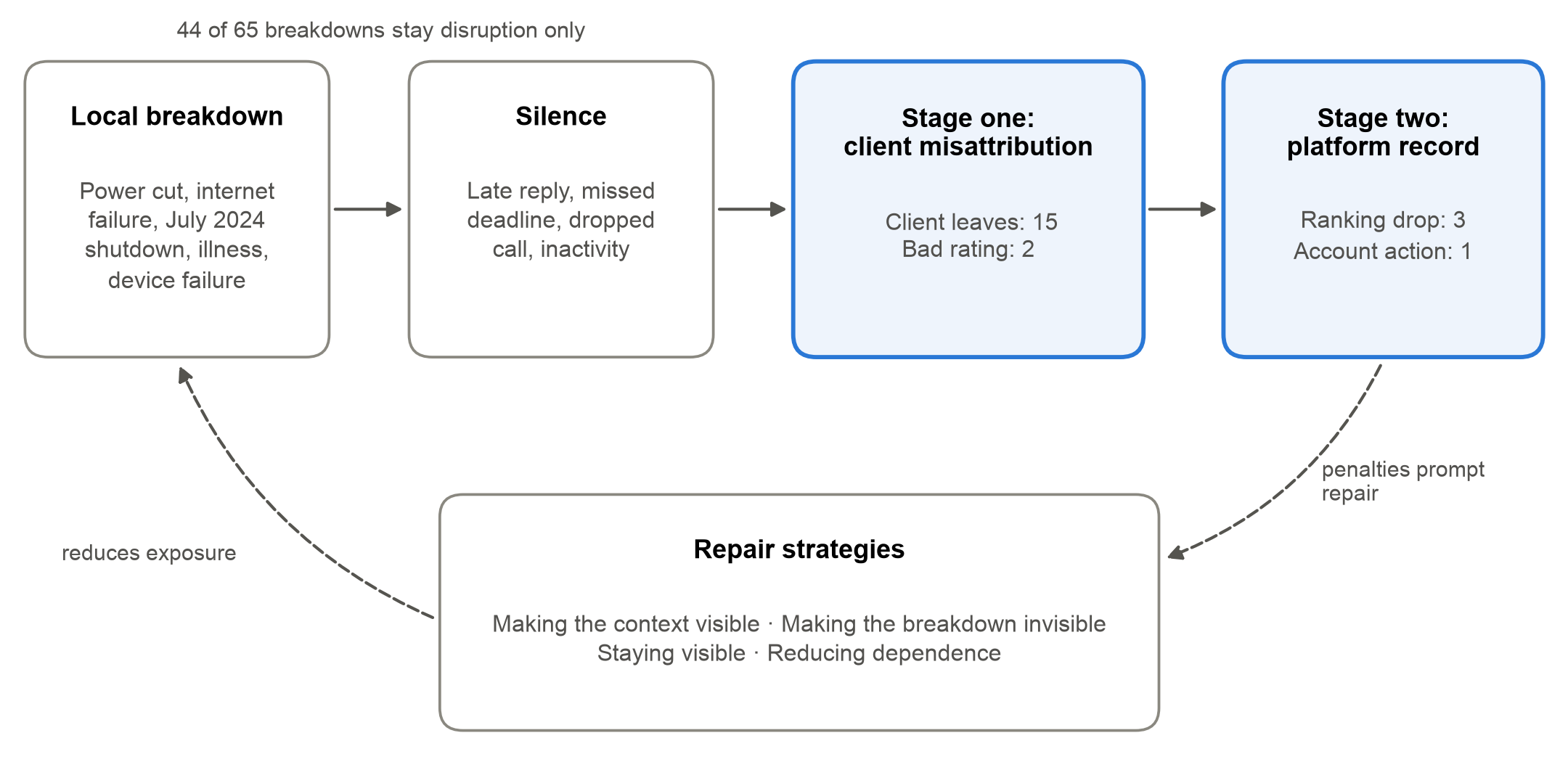}
\caption{Client-mediated misattribution. Numbers are breakdown incidents. Dashed arrows show how repair responds to penalties and reduces exposure.}\label{fig:4}
\end{figure}

The findings answer the first research question with a two-stage account. In the first stage, a client who cannot see the worker's situation attributes an infrastructural failure to the worker's character or effort. In the second stage, a platform reputation system records that judgment without its context and carries it forward. We call this process client-mediated misattribution. \hyperref[fig:4]{Figure~4} shows the process, including the repair strategies that feed back into it.

The process depends on three conditions, each visible in the data. First, the platform evaluates workers through traces of responsiveness and delivery, which depend on infrastructure. Second, the evaluator is distant and cannot see local conditions. Third, the platform stores the evaluation without any record of its cause and without a practical way for the worker to contest it. Where any of these conditions is absent, the process weakens. In our data, local clients who could be phoned, and foreign clients who had seen the news, did not penalize the worker.

\subsection{Contributions}\label{contributions}

The study makes three contributions. First, it adds to research on algorithmic management and platform reputation systems. That research locates unfairness in opaque algorithms (\textcolor{themeaccent}{Rahman, 2021}) or in customer bias (\textcolor{themeaccent}{Rosenblat et al., 2017}). This study identifies a third source: an honest but mistaken client judgment that the platform stores. The platform does not need to misread the worker itself. It needs only to keep the client's judgment and remove its context. Opacity also runs in two directions. Workers cannot see how the platform scores them, and the people who score them cannot see the workers' conditions.

Second, the study extends research on attribution in distributed work. Cramton (2001) and Cramton et al. (2007) showed that remote partners blame the person when the situation is invisible. Their participants worked in teams that shared a future and could correct impressions over time. In platform markets, the relationship is often a single transaction, the client can replace the worker immediately, and the misattribution becomes a public score that later clients will see. The same cognitive bias is therefore likely to have larger and longer-lasting consequences.

Third, the study contributes to research on information systems in development. Star and Ruhleder (1996) observed that infrastructure becomes visible when it breaks down. Our data suggest that breakdown is visible only to some of the people it affects. For the worker, the breakdown is impossible to miss. For the client and the platform, it is silence. That gap moves the cost of national infrastructural failure onto individual workers. It also adds a mechanism to the idea of adverse digital incorporation (\textcolor{themeaccent}{Heeks, 2022}): workers are included in global markets on terms that count their country's infrastructural failures against them personally.

\subsection{Implications for practice}\label{implications-for-practice}

The findings suggest three changes to platform design. Each would need testing, because we had no access to platform data. First, platforms could show context to clients. A notice that a freelancer's region is affected by a documented outage could address the first stage directly. The cases in this study suggest that clients who understood the situation did not penalize the worker. Second, platforms could protect records during verified regional outages, for example by suspending response-rate calculations and delaying review windows. Fiverr already excludes unresponsive clients from freelancers' scores (\textcolor{themeaccent}{Fiverr, n.d.-c}), so a similar rule for regional outages would extend an existing principle. Third, platforms could let workers switch on availability settings after an outage has started, or without internet access. The fair-management principle of the Fairwork framework, which calls for due process in decisions that affect workers, gives a standard for such changes (\textcolor{themeaccent}{Heeks et al., 2021}).

For governments, the findings suggest that internet shutdowns impose costs on individual digital workers that can continue after connectivity returns. Programs that train freelancers could also support reliable power and connectivity, since the workers in this study paid for these themselves.

\subsection{Limitations and future research}\label{limitations-and-future-research}

This study has limitations. It is a secondary analysis of interviews that were not designed for this question. The guide did not ask participants what happened to their ratings or rankings after outages, which is one reason recovery was often unclear and the platform's role may be understated. The accounts are retrospective. The sample is young, relies heavily on Fiverr, and includes a cluster of participants from one district, recruited through referral. Location type was inferred from the interviews. The client side of the process is visible only through workers' accounts.

Future research could address these gaps in three ways. Interviews or experiments with clients could test how they interpret a freelancer's silence with and without information about an outage. An analysis of public profile metrics before and after the July 2024 shutdown could measure the platform stage directly. Comparisons with other countries that experience frequent shutdowns could show whether the process takes the same form elsewhere.

\section{Conclusion}\label{conclusion}

Meem Jahan's two loyal clients had seen the news. Her other five had not, and they moved their projects to someone else. The question in the title has a clear answer in these data: when the power or the internet fails, the worker is usually blamed. This paper has argued that her experience reflects a wider process. Distant clients misread silence caused by infrastructural breakdown, and platform reputation systems record that misreading as a lasting judgment. Online freelancers in Bangladesh absorb most breakdowns, but the ones that matter strike at moments when a client is waiting and cannot see why. Workers respond by explaining, by buying their own backup infrastructure, by staying constantly visible, and by spreading their risk. Their ability to do so depends on money and on how far away their clients are. Online labor platforms promise to judge workers by their skill and reliability. As long as their reputation systems record silence without its cause, they will also judge workers by the reliability of their country's power grid.

\section*{Declarations}\addcontentsline{toc}{section}{Declarations}\label{declarations}

\textbf{Author contributions.} \textbf{Naim Bin Hasan:} Conceptualization, Methodology, Investigation, Data curation, Formal analysis, Validation, Funding acquisition, Writing -- original draft, Writing -- review and editing. \textbf{Fahim Bin Hasan:} Validation (verification of the incident extraction, the AI-assisted coding and the reliability check against the transcripts), Writing -- review and editing.

\textbf{Use of generative AI.} Claude (Anthropic), a large language model, assisted with the first extraction of incidents from the transcripts. A GPT model (OpenAI) served as the blind second coder in the reliability check. The authors verified every incident against the original transcripts, made all analytic and interpretive decisions, checked every reference, and take full responsibility for the content.

\textbf{Ethics.} The study was approved by the Institutional Review Board of Florida Atlantic University (protocol IRB2502046). Participants gave verbal consent. All names are pseudonyms.

\textbf{Data availability.} The interview transcripts contain identifying information and cannot be shared. The coding scheme is available from the corresponding author on request.

\textbf{Funding.} Data collection was supported by the American Institute of Bangladesh Studies (AIBS) Fellowship and the Florida Atlantic University Graduate Thesis and Dissertation Scholarship.

\textbf{Acknowledgments.} We thank the Bangladeshi freelancers who shared their time and experiences, and Naim Bin Hasan's advisers at Florida Atlantic University for their guidance.

\textbf{Declaration of interest.} The authors have no competing interests to declare.

\section*{References}\addcontentsline{toc}{section}{References}\label{references}
\begingroup\small\setlength{\parindent}{-0.4in}\setlength{\leftskip}{0.4in}\setlength{\parskip}{4pt}\raggedright

Agrawal, A., Lacetera, N., \& Lyons, E. (2016). Does standardized information in online markets disproportionately benefit job applicants from less developed countries? \emph{Journal of International Economics, 103}, 1--12. https://doi.org/10.1016/j.jinteco.2016.08.003

Allcott, H., Collard-Wexler, A., \& O'Connell, S. D. (2016). How do electricity shortages affect industry? Evidence from India. \emph{American Economic Review, 106}(3), 587--624. https://doi.org/10.1257/aer.20140389

Alter, S. (2014). Theory of workarounds. \emph{Communications of the Association for Information Systems, 34}, Article 55. https://doi.org/10.17705/1CAIS.03455

Anwar, M. A., \& Graham, M. (2021). Between a rock and a hard place: Freedom, flexibility, precarity and vulnerability in the gig economy in Africa. \emph{Competition \& Change, 25}(2), 237--258. https://doi.org/10.1177/1024529420914473

Avgerou, C. (2008). Information systems in developing countries: A critical research review. \emph{Journal of Information Technology, 23}(3), 133--146. https://doi.org/10.1057/palgrave.jit.2000136

Azad, B., \& King, N. (2012). Institutionalized computer workaround practices in a Mediterranean country: An examination of two organizations. \emph{European Journal of Information Systems, 21}(4), 358--372. https://doi.org/10.1057/ejis.2011.48

Beerepoot, N., \& Lambregts, B. (2015). Competition in online job marketplaces: Towards a global labour market for outsourcing services? \emph{Global Networks, 15}(2), 236--255. https://doi.org/10.1111/glob.12051

Belson, D. (2024, August 1). \emph{A recent spate of Internet disruptions: July 2024}. The Cloudflare Blog. https://blog.cloudflare.com/a-recent-spate-of-internet-disruptions-july-2024

Billah, M., \& Hossain, M. (2024, July 31). The many nightmares of Bangladeshi freelancers. \emph{The Business Standard}. https://www.tbsnews.net/node/906691

Billah, M., \& Khan, T. H. (2026). From flexibility to fragility: Exploring the experiences of gig workers in Bangladesh. \emph{Cogent Social Sciences, 12}(1). https://doi.org/10.1080/23311886.2026.2648880

Bucher, E. L., Schou, P. K., \& Waldkirch, M. (2021). Pacifying the algorithm: Anticipatory compliance in the face of algorithmic management in the gig economy. \emph{Organization, 28}(1), 44--67. https://doi.org/10.1177/1350508420961531

Butterfield, L. D., Borgen, W. A., Amundson, N. E., \& Maglio, A.-S. T. (2005). Fifty years of the critical incident technique: 1954--2004 and beyond. \emph{Qualitative Research, 5}(4), 475--497. https://doi.org/10.1177/1468794105056924

Cameron, L. D., \& Rahman, H. (2022). Expanding the locus of resistance: Understanding the co-constitution of control and resistance in the gig economy. \emph{Organization Science, 33}(1), 38--58. https://doi.org/10.1287/orsc.2021.1557

Cohen, J. (1960). A coefficient of agreement for nominal scales. \emph{Educational and Psychological Measurement, 20}(1), 37--46. https://doi.org/10.1177/001316446002000104

Cramton, C. D. (2001). The mutual knowledge problem and its consequences for dispersed collaboration. \emph{Organization Science, 12}(3), 346--371. https://doi.org/10.1287/orsc.12.3.346.10098

Cramton, C. D., Orvis, K. L., \& Wilson, J. M. (2007). Situation invisibility and attribution in distributed collaborations. \emph{Journal of Management, 33}(4), 525--546. https://doi.org/10.1177/0149206307302549

Curchod, C., Patriotta, G., Cohen, L., \& Neysen, N. (2020). Working for an algorithm: Power asymmetries and agency in online work settings. \emph{Administrative Science Quarterly, 65}(3), 644--676. https://doi.org/10.1177/0001839219867024

Filippas, A., Horton, J. J., \& Golden, J. M. (2022). Reputation inflation. \emph{Marketing Science, 41}(4), 733--745. https://doi.org/10.1287/mksc.2022.1350

Fiverr. (n.d.-a). \emph{Everything you need to know about response time and rate}. Fiverr Help Center. Retrieved October 4, 2026, from https://help.fiverr.com/hc/en-us/articles/360011451678-Everything-you-need-to-know-about-response-time-and-rate

Fiverr. (n.d.-b). \emph{Setting availability for freelancers}. Fiverr Help Center. Retrieved October 4, 2026, from https://help.fiverr.com/hc/en-us/articles/360015529197-Setting-Availability-for-freelancers

Fiverr. (n.d.-c). \emph{Success score}. Fiverr Help Center. Retrieved October 4, 2026, from https://help.fiverr.com/hc/en-us/articles/21965360854673-Success-score

Fiverr. (n.d.-d). \emph{Understanding Fiverr's freelancer levels}. Fiverr Help Center. Retrieved October 4, 2026, from https://help.fiverr.com/hc/en-us/articles/360010560118-Understanding-Fiverr-s-freelancer-levels

Flanagan, J. C. (1954). The critical incident technique. \emph{Psychological Bulletin, 51}(4), 327--358. https://doi.org/10.1037/h0061470

Freyburg, T., \& Garbe, L. (2018). Blocking the bottleneck: Internet shutdowns and ownership at election times in sub-Saharan Africa. \emph{International Journal of Communication, 12}, 3896--3916. https://ijoc.org/index.php/ijoc/article/view/8546

Fuller, L., \& Smith, V. (1991). Consumers' reports: Management by customers in a changing economy. \emph{Work, Employment and Society, 5}(1), 1--16. https://doi.org/10.1177/0950017091005001002

Gilbert, D. T., \& Malone, P. S. (1995). The correspondence bias. \emph{Psychological Bulletin, 117}(1), 21--38. https://doi.org/10.1037/0033-2909.117.1.21

Graham, M., Hjorth, I., \& Lehdonvirta, V. (2017). Digital labour and development: Impacts of global digital labour platforms and the gig economy on worker livelihoods. \emph{Transfer: European Review of Labour and Research, 23}(2), 135--162. https://doi.org/10.1177/1024258916687250

Graham, S., \& Thrift, N. (2007). Out of order: Understanding repair and maintenance. \emph{Theory, Culture \& Society, 24}(3), 1--25. https://doi.org/10.1177/0263276407075954

Heeks, R. (2022). Digital inequality beyond the digital divide: Conceptualizing adverse digital incorporation in the global South. \emph{Information Technology for Development, 28}(4), 688--704. https://doi.org/10.1080/02681102.2022.2068492

Heeks, R., Graham, M., Mungai, P., Van Belle, J.-P., \& Woodcock, J. (2021). Systematic evaluation of gig work against decent work standards: The development and application of the Fairwork framework. \emph{The Information Society, 37}(5), 267--286. https://doi.org/10.1080/01972243.2021.1942356

Hinds, P. J., \& Mortensen, M. (2005). Understanding conflict in geographically distributed teams: The moderating effects of shared identity, shared context, and spontaneous communication. \emph{Organization Science, 16}(3), 290--307. https://doi.org/10.1287/orsc.1050.0122

Internet Society Pulse. (2024). \emph{Internet suspended in Bangladesh (18 July 2024)} {[}Shutdown tracker entry{]}. https://pulse.internetsociety.org/en/shutdowns/internet-suspended-in-bangladesh-18-july-2024/

Jarvenpaa, S. L., \& Leidner, D. E. (1999). Communication and trust in global virtual teams. \emph{Organization Science, 10}(6), 791--815. https://doi.org/10.1287/orsc.10.6.791

Kelley, H. H. (1973). The processes of causal attribution. \emph{American Psychologist, 28}(2), 107--128. https://doi.org/10.1037/h0034225

Kelley, H. H., \& Michela, J. L. (1980). Attribution theory and research. \emph{Annual Review of Psychology, 31}, 457--501. https://doi.org/10.1146/annurev.ps.31.020180.002325

Kellogg, K. C., Valentine, M. A., \& Christin, A. (2020). Algorithms at work: The new contested terrain of control. \emph{Academy of Management Annals, 14}(1), 366--410. https://doi.org/10.5465/annals.2018.0174

Klein, H. K., \& Myers, M. D. (1999). A set of principles for conducting and evaluating interpretive field studies in information systems. \emph{MIS Quarterly, 23}(1), 67--93. https://doi.org/10.2307/249410

Kokkodis, M., \& Ipeirotis, P. G. (2016). Reputation transferability in online labor markets. \emph{Management Science, 62}(6), 1687--1706. https://doi.org/10.1287/mnsc.2015.2217

Landis, J. R., \& Koch, G. G. (1977). The measurement of observer agreement for categorical data. \emph{Biometrics, 33}(1), 159--174. https://doi.org/10.2307/2529310

Larkin, B. (2013). The politics and poetics of infrastructure. \emph{Annual Review of Anthropology, 42}, 327--343. https://doi.org/10.1146/annurev-anthro-092412-155522

Lehdonvirta, V., Kässi, O., Hjorth, I., Barnard, H., \& Graham, M. (2019). The global platform economy: A new offshoring institution enabling emerging-economy microproviders. \emph{Journal of Management, 45}(2), 567--599. https://doi.org/10.1177/0149206318786781

Mazmanian, M., Orlikowski, W. J., \& Yates, J. (2013). The autonomy paradox: The implications of mobile email devices for knowledge professionals. \emph{Organization Science, 24}(5), 1337--1357. https://doi.org/10.1287/orsc.1120.0806

Möhlmann, M., Zalmanson, L., Henfridsson, O., \& Gregory, R. W. (2021). Algorithmic management of work on online labor platforms: When matching meets control. \emph{MIS Quarterly, 45}(4), 1999--2022. https://doi.org/10.25300/MISQ/2021/15333

Moreno, A., \& Terwiesch, C. (2014). Doing business with strangers: Reputation in online service marketplaces. \emph{Information Systems Research, 25}(4), 865--886. https://doi.org/10.1287/isre.2014.0549

Office of the United Nations High Commissioner for Human Rights. (2025). \emph{OHCHR fact-finding report: Human rights violations and abuses related to the protests of July and August 2024 in Bangladesh}. https://www.ohchr.org/en/documents/country-reports/ohchr-fact-finding-report-human-rights-violations-and-abuses-related

Pallais, A. (2014). Inefficient hiring in entry-level labor markets. \emph{American Economic Review, 104}(11), 3565--3599. https://doi.org/10.1257/aer.104.11.3565

Rahman, H. A. (2021). The invisible cage: Workers' reactivity to opaque algorithmic evaluations. \emph{Administrative Science Quarterly, 66}(4), 945--988. https://doi.org/10.1177/00018392211010118

Rani, U., \& Furrer, M. (2021). Digital labour platforms and new forms of flexible work in developing countries: Algorithmic management of work and workers. \emph{Competition \& Change, 25}(2), 212--236. https://doi.org/10.1177/1024529420905187

Rosenblat, A., Levy, K. E. C., Barocas, S., \& Hwang, T. (2017). Discriminating tastes: Uber's customer ratings as vehicles for workplace discrimination. \emph{Policy \& Internet, 9}(3), 256--279. https://doi.org/10.1002/poi3.153

Rosenblat, A., \& Stark, L. (2016). Algorithmic labor and information asymmetries: A case study of Uber's drivers. \emph{International Journal of Communication, 10}, 3758--3784. https://ijoc.org/index.php/ijoc/article/view/4892

Star, S. L. (1999). The ethnography of infrastructure. \emph{American Behavioral Scientist, 43}(3), 377--391. https://doi.org/10.1177/00027649921955326

Star, S. L., \& Ruhleder, K. (1996). Steps toward an ecology of infrastructure: Design and access for large information spaces. \emph{Information Systems Research, 7}(1), 111--134. https://doi.org/10.1287/isre.7.1.111

Stephany, F., Kässi, O., Rani, U., \& Lehdonvirta, V. (2021). Online Labour Index 2020: New ways to measure the world's remote freelancing market. \emph{Big Data \& Society, 8}(2). https://doi.org/10.1177/20539517211043240

Sutherland, W., Jarrahi, M. H., Dunn, M., \& Nelson, S. B. (2020). Work precarity and gig literacies in online freelancing. \emph{Work, Employment and Society, 34}(3), 457--475. https://doi.org/10.1177/0950017019886511

Upwork. (n.d.-a). \emph{All about your Job Success Score}. Upwork Help Center. Retrieved October 4, 2026, from https://support.upwork.com/hc/en-us/articles/211068358-All-about-your-Job-Success-Score

Upwork. (n.d.-b). \emph{How is my Job Success Score calculated?} Upwork Help Center. Retrieved October 4, 2026, from https://support.upwork.com/hc/en-us/articles/38437458199059-How-is-my-Job-Success-Score-calculated

Wagner, B. (2018). Understanding internet shutdowns: A case study from Pakistan. \emph{International Journal of Communication, 12}, 3917--3938. https://ijoc.org/index.php/ijoc/article/view/8545

Walsham, G. (2006). Doing interpretive research. \emph{European Journal of Information Systems, 15}(3), 320--330. https://doi.org/10.1057/palgrave.ejis.3000589

Walsham, G., \& Sahay, S. (2006). Research on information systems in developing countries: Current landscape and future prospects. \emph{Information Technology for Development, 12}(1), 7--24. https://doi.org/10.1002/itdj.20020

Walther, J. B. (1996). Computer-mediated communication: Impersonal, interpersonal, and hyperpersonal interaction. \emph{Communication Research, 23}(1), 3--43. https://doi.org/10.1177/009365096023001001

Wood, A. J., Graham, M., Lehdonvirta, V., \& Hjorth, I. (2019). Good gig, bad gig: Autonomy and algorithmic control in the global gig economy. \emph{Work, Employment and Society, 33}(1), 56--75. https://doi.org/10.1177/0950017018785616

Wood, A. J., \& Lehdonvirta, V. (2023). Platforms disrupting reputation: Precarity and recognition struggles in the remote gig economy. \emph{Sociology, 57}(5), 999--1016. https://doi.org/10.1177/00380385221126804
\endgroup

\end{document}